\documentclass[10pt,notitlepage]{article}   

\usepackage{color,graphicx,bm,amsmath}

\begin{document}

\title{ Role of incoherent substrate reflections in photonic crystal 
spectroscopy}

\author{D.~M.~Whittaker and  G.~A.~Gehring \\Department of Physics and
Astronomy, University of Sheffield,\\ Sheffield S3 7RH, UK.}

\date{\today}

\maketitle

\begin{abstract}
A method is presented for modelling the optical properties of a
photonic crystal structure mounted on a substrate which is thick
enough that the light reflected from the back is incoherent with
reflections from the front. Transmission and reflection spectra are
presented for the cases where the structure is a multilayer planar
microcavity and an opal photonic crystal. The contributions from the
incoherent reflections can be very large, and in the photonic crystal
case, change the positions of the spectral peaks.\\
\end{abstract}

There are many situations where a photonic crystal structure which is
studied experimentally is mounted on a thick substrate. In
modelling the optical properties of the actual structure, we need to
consider the effects of interference between light reflected off
different interfaces. However, if we also allow interference between
waves reflected from the structure and off back of the substrate, the
spectra which are obtained are very unrealistic, being dominated by
closely spaced Fabry-Perot fringes. This problem is typically avoided
in calculations by treating the substrate as semi-infinite,
eliminating any reflections.  In reality, the interference effects are
not seen because inhomogeneities in the substrate, such as thickness
variations, mean that any waves which has passed through the substrate
is effectively incoherent with the front reflections. In this letter,
we present a simple but much more realistic theoretical treatment
which incorporates multiple reflections in the substrate as incoherent
waves. Our results show that these reflections can make very
significant contributions to the spectra, so the semi-infinite
substrate approach is not appropriate for accurate comparisons with
experimental data.

The method we describe is very adaptable, and can be applied to any
structure for which the intensity transmittance and reflectance on a
semi-infinite substrate can be calculated, either analytically or
numerically. As indicated in Fig.(1), we call these quantities $T_f$ and
$R_f$ for light passing from air into the substrate, $\tilde{T}_f$ and
$\tilde{R}_f$ for the reverse direction. We also require the the
transmittance and reflectance of the back of the substrate, $T_b$ and
$R_b$, which come from the standard Fresnel expressions.

\begin{figure}
\begin{center}
\includegraphics[scale=.3]{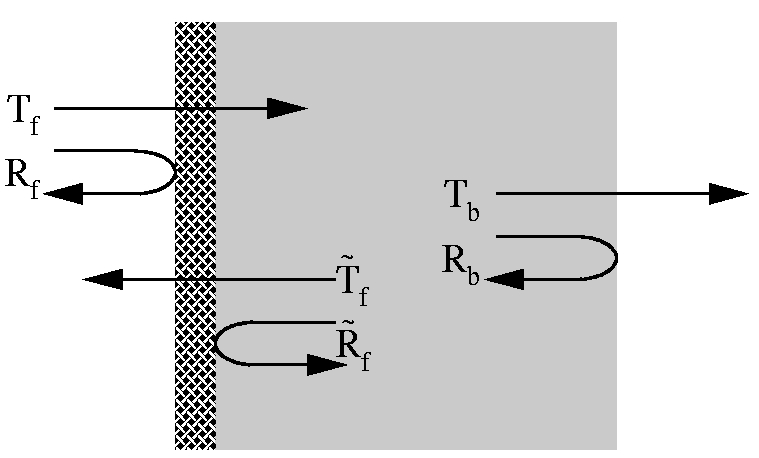}
\end{center}
\caption{Definitions of the front and back 
transmittance and reflectance coefficients.}
\end{figure}

The total transmittance, ${\cal T}$, is obtained by summing over an
infinite number of incoherent reflections from the surfaces of the
substrate.  If the absorption coefficient of the substrate, $\alpha$, is
non-zero there is an attenuation $\exp{(-\alpha l)}$ on every pass.
The result is
\begin{align}
{\cal T} & =
T_f 
\left[
1+R_b \tilde{R}_f \, e^{-2 \alpha l}+(R_b \tilde{R}_f \, e^{-2 \alpha l})^2
+ \cdots
\right] T_b \, e^{-\alpha l}
\nonumber \\
&=
T_f \, (1-e^{-2 \alpha l} R_b \tilde{R}_f)^{-1} T_b \, e^{-\alpha l}  
\end{align}
Similarly the total reflectance, ${\cal R}$, is
\begin{align}
{\cal R} &= R_f + T_f \left[
1+R_b \tilde{R}_f \, e^{-2 \alpha l} \right.
\nonumber \\
&\quad \left.  + (R_b \tilde{R}_f \, e^{-2 \alpha l})^2 + \cdots
\right] R_b \tilde{T}_f e^{-2 \alpha l}  
\nonumber \\
&= R_f+ T_f \, (1-e^{-2 \alpha l} R_b \tilde{R}_f)^{-1} R_b
\tilde{T}_f\, e^{-2\alpha l}.
\end{align}
Since there is no interference, these results are independent of the
thickness of the substrate, $l$, except for the attenuation due to
absorption.

For a transparent substrate, with no structure on the surface, the
reflectances are all the same, $R_f=\tilde{R}_f=R_b=R$, and also
$T_f=\tilde{T}_f=T_b=1-R$.  Eqs.(1) and (2) then reproduce the well
known results for incoherent reflections,
\begin{align}
{\cal T}=\frac{1-R}{1+R}
\qquad
{\cal R}=\frac{2 R}{1+R}=1-\cal{T}.
\end{align}
A further check is the case of a thin film on the surface of the
substrate. The transmittance of such a film with incoherent substrate
reflections is given by Swanepoel\cite{swanepoel}. It is straight
forward to obtain expressions for $T_f$ and $\tilde{R}_f$ for a simple
film, and Eq.(1) then correctly reproduces this
result\cite{derivation}.

\begin{figure}[t]
\begin{center}
\includegraphics[scale=.3]{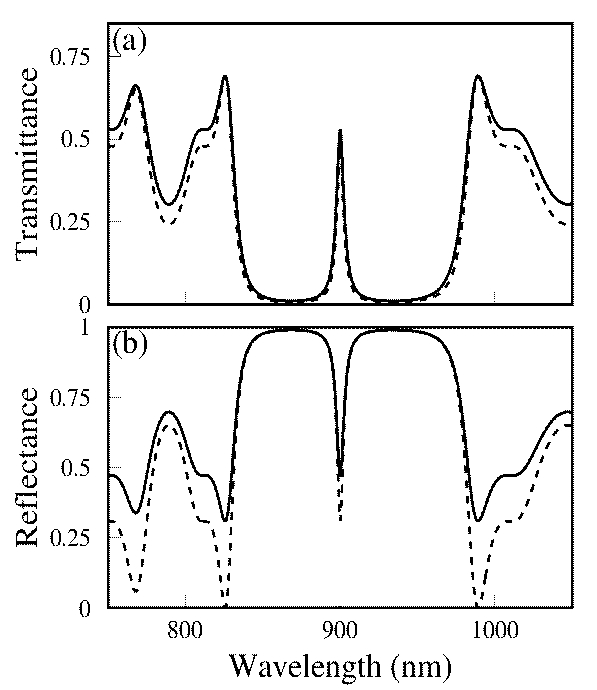}
\end{center}
\caption{ Solid lines: normal incidence transmission (a) and
reflection (b) spectra for the microcavity structure on a 
substrate. Dashed lines: results obtained when the
 incoherent multiple reflections in the substrate are neglected.}
\end{figure}

Turning to the case of a one-dimensional photonic crystal, we next
calculate the properties of a planar microcavity on a thick substrate.
The parameters used are typical for a GaAs/AlGaAs structure, with a
design wavelength of $\lambda_0=900$nm. The cavity is a $\lambda/2$
thickness\cite{thickness} layer of refractive index 3.5.  with the
mirrors on either side comprising 10 bilayer Bragg stacks with layer
thicknesses $\lambda/4$ and high and low indicies of 3.5 and 3. The
substrate index is 3.5 and its absorption coefficient $\alpha=0$.

In this case, it is easiest to obtain the front coefficients, $R_f$,
$\tilde{R}_f$, $T_f$ and $\tilde{T}_f$ from a transfer matrix
calculation\cite{transfer}. We have done this numerically, and the
transmission and reflection spectra calculated using Eqs.(1) and (2)
are shown in Fig.(2). The dashed lines on the figure show the results
obtained when multiple reflections in the substrate are
ignored, by putting $R_b=0$; for the reflectance, this is simply
$R_f$, the structure reflectance, while for the transmittance it is
$T_f T_b$. The incoherent reflections always add to ${\cal T}$ and ${\cal
R}$, and their contribution is significant, except where the
reflectance of the structure is high, and no light passes through the
microcavity into the substrate. Where the structure reflectance is zero
or very small, all the light passes through in both directions, and we
just see the reflectance and transmittance of the back of the
substrate, which are $\sim 0.31$ and 0.69 respectively.

\begin{figure}[t]
\begin{center}
\includegraphics[scale=.3]{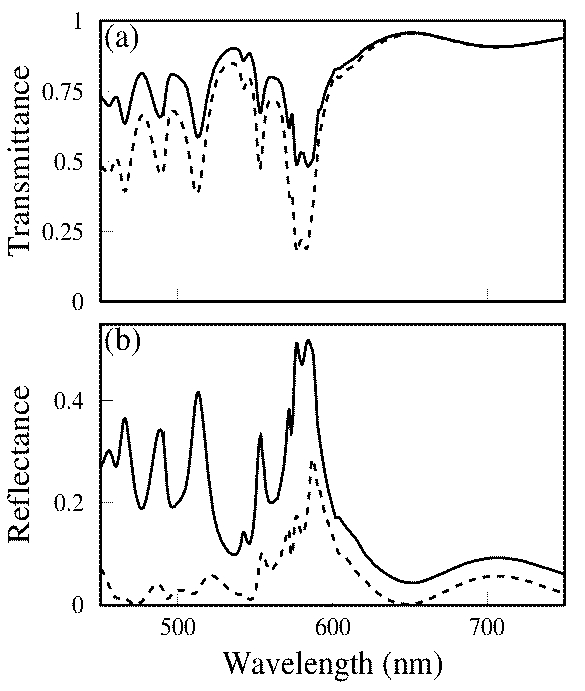}
\end{center}
\caption{Normal incidence, unpolarised transmission (a) and reflection
(b) spectra for the 3-layer opal structure on a glass substrate. The
solid and dashed lines are results with, and without, incoherent
reflections, as in Fig.(2).
}
\end{figure}

The situation becomes more complicated when we consider a two or three
dimensional photonic crystal fabricated on a substrate, so the
structure has an in-plane periodicity.  The effect of this periodicity
is to give rise to diffraction, so for some wavelengths a single incident
wave will produce multiple diffracted waves in the
substrate, or in both substrate and air. Despite this, Eqs.(1) and (2)
still apply, though the symbols needs to be
reinterpreted as matrices.

Consider a wavelength at which there are $N_a$ waves in air, and $N_s$ waves
in the substrate, including both straight through and diffracted
directions. In general, photonic crystals cause mixing of polarisations,
so both TE and TM waves need to be included and the $N$ values will be 
twice the number of directions.
We can then describe the front reflection and transmission
processes by matrices; for example, $T_f$ becomes a $N_s \times N_a$\,
matrix, whose $(i,j)$ element gives the intensity generated in
substrate wave $i$ for illumination in exterior wave $j$. We assume
again that there is some method available to obtain these matrices; in
the example presented in Fig.(3), they are calculated using the
scattering matrix method of Ref.\cite{scatter}. The back
coefficients $T_b$ and $R_b$ become diagonal matrices, with elements
given by the standard Fresnel expressions for finite angle
transmittance and reflectance. Knowing these matrix elements, we can
add up intensities, as in the derivation of Eqs.(1) and (2), with the
complication that we now need to sum over every possible path,
comprising the combinations of the different directions and
polarisations that can occur on each pass through the substrate.  This
summation is obtained by replacing expressions such as $R_b
\tilde{R}_f$ by the equivalent matrix products, since the sum over the
internal index in the product replicates the sum over paths.  Eqs.(1)
and (2) then give the total transmittance and reflectance ${\cal T}$ and
${\cal R}$, now as $N_a \times N_a$ matrices.

In Fig.(3), we show the results of this procedure for an opal structure
on a glass substrate. The opal consists of three hcp-ordered layers of
500nm spheres, with refractive index 1.5; the substrate also has
index 1.5 and absorption $\alpha=0$. 
For wavelengths longer than 649.5nm, there is no diffraction, and the
physics is similar to the planar microcavity, though we have $2 \times
2$ matrices because of the polarisation mixing induced by the photonic
crystal. The structure reflectance is very low, so the substrate
contributes just one back reflection, which adds a constant $0.04$ to
the reflectance but does not affect the transmittance.

At shorter wavelengths, there are six diffracted directions in the substrate,
but, in the range of the figure, none in air, so $N_a=2$ and
$N_s=14$. Again, the structure reflectance is quite low, but the
diffracted waves, unable to propagate in air, experience total
internal reflection at the back of the substrate. They then interact
with the opal again and are partially re-diffracted back into the
normal direction, to provide strong contributions to both the
reflectance and the transmittance. It is notable in the reflectance
that these contributions not only are much larger than those from the
front reflection, but they are also different spectrally,
with peaks appearing at different wavelengths. It is clear
from this that it may be very misleading to neglect the substrate
reflections, though we believe that this has been done in all previous
published calculations.

In conclusion we have shown that, if transmission and reflection
spectra can be calculated for an optical structure, there is a very
simple formulism by which to add the contributions due to multiple
incoherent reflections occuring when the structure is mounted on a
finite substrate. We have used this method to calculate the optical
properties of a planar microcavity structure and an opal photonic
crystal, and shown that the substrate reflections can modify very
significantly the spectra which are obtained.

\appendix

\section*{Supplementary Information:
Derivation of transmission of thin film on a substrate}

\makeatletter 
\renewcommand{\thefigure}{S\@arabic\c@figure}
\renewcommand{\thetable}{S\@arabic\c@table}
\renewcommand{\theequation}{S\@arabic\c@equation}
\makeatother

\setcounter{figure}{0}
\setcounter{equation}{0}
\setcounter{table}{0}

In this section, we calculate the total transmission of a thin film of
thickness $d$ and refractive index $n-ik$ on a transparent substrate
with index $s$. We show that our result agrees with the expression
given in Eq.(4) of Ref.\cite{swanepoel}. The derivation is for the
case where the film is weakly absorbing, so there is attenuation of
light passing through the film (given by the attenuation factor, $x$),
but the imaginary part of the refractive index, $k$, is neglected in
the expressions for the reflection and transmission coefficients at
the interfaces.

\begin{figure}[here]

\begin{center}
\includegraphics[scale=.35]{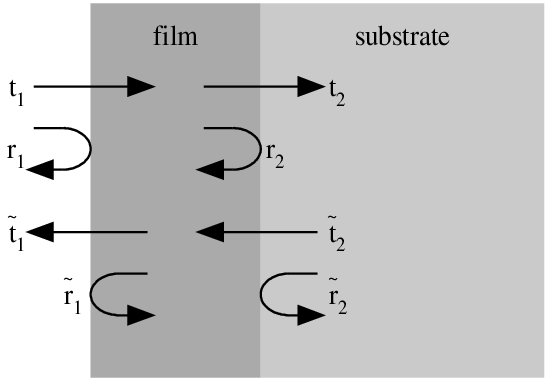}
\end{center}
\caption{Definitions of the 
amplitude reflection and transmission
coefficients for the film calculation.}
\end{figure}

We first evaluate $T_f$ and $\tilde{R}_f$ for the thin film. They are
obtained, in the usual way, using a similar approach to our derivation
of Eqs.(1) and (2), except adding amplitudes because the reflections
are coherent. On each double pass through the film, there is a phase
change $\phi$ and an amplitude reduction, $x$, which are given by
\begin{align}
\phi=\frac{4 \pi n d}{\lambda} \qquad
x=\exp{\left(- \frac{4\pi k d}{\lambda} \right)}=\exp{(-\alpha d)}.
\end{align}
The transmittance of the film is then
\begin{align}
T_f=\frac{|t_1 t_2|^2 x}{s \, |1-r_2 \tilde{r}_1 x e^{i \phi}|^2},
\end{align}
where the amplitude reflection and transmission coefficients at
the interfaces, $t_1$, $\tilde{t}_1$, $t_2$, $\tilde{t}_2$, $r_1$,
$\tilde{r}_1$, $r_2$ and $\tilde{r}_2$, are defined in Fig.(S1). 
Using the the Fresnel formulae, they are
\begin{align}
t_1&=\frac{2n}{n+1}, \qquad \tilde{t}_1=\frac{2}{n+1}, 
\qquad t_2=\frac{2s}{n+s}, \qquad
\tilde{t}_2=\frac{2n}{n+s},
\nonumber \\
r_1&=\frac{n-1}{n+1}=-\tilde{r_1}, \qquad
r_2=\frac{s-n}{s+n}=-\tilde{r}_2.
\end{align}
Substituting these expressions into Eq.(S2), we obtain
\begin{align}
 T_f&= 
%\frac{
\frac{1}{s} 
\left(\frac{2n}{n+1}\right)^2
\left(\frac{2s}{n+s}\right)^2
x
%}
\nonumber \\
%{
&\times
\left[
1-2x \left(\frac{1-n}{1+n}\right)  \left(\frac{s-n}{s+n}\right) \cos{\phi}
+x^2 \left(\frac{1-n}{1+n}\right)^2  \left(\frac{s-n}{s+n}\right)^2 
\right]^{-1}
%}
\nonumber \\
&=\frac{16 n^2 s x}{\Delta},
\end{align}
where the denominator is
\begin{align}
\Delta=
{(n+1)^2 (n+s)^2 - 2x (n^2-1)(n^2-s^2)\cos{\phi}
+x^2 (n-1)^2 (n-s)^2}.
\end{align}
The reflectance of the film from the substrate side is
\begin{align}
\tilde{R}_f&= 
\left|
\tilde{r}_2+
\frac{\tilde{r}_1 t_2 \tilde{t}_2 x e^{i \phi}}
{1-r_2 \tilde{r}_1 x e^{i \phi}}
\right|^2
=
\left|
\frac{
\tilde{r}_2-(\tilde{r}_2 r_2 - \tilde{t}_2 t_2) \tilde{r}_1 x e^{i \phi}}
{1-r_2 \tilde{r}_1 x e^{i \phi}}
\right|^2
%\nonumber \\
=
\left|
\frac{
\tilde{r}_2+\tilde{r}_1 x e^{i \phi}}
{1-r_2 \tilde{r}_1 x e^{i \phi}}
\right|^2,
\end{align}
where we have used the identity
\begin{align}
\tilde{t}_2 t_2 - \tilde{r}_2 r_2
=
\frac{4 sn}{(n+s)^2}
+\frac{(n-s)^2}{(n+s)^2}=1.
\end{align}
Substituting from Eq.(S3), we get
\begin{align}
\tilde{R}_f
&=\left[
\left(\frac{n-s}{n+s}\right)^2
-2x \left(\frac{1-n}{1+n}\right)  \left(\frac{s-n}{s+n}\right) \cos{\phi}
+x^2 \left(\frac{1-n}{1+n}\right)^2   
\right] 
\nonumber \\
&\times
\left[
1-2x \left(\frac{1-n}{1+n}\right)  \left(\frac{s-n}{s+n}\right) \cos{\phi}
+x^2 \left(\frac{1-n}{1+n}\right)^2  \left(\frac{s-n}{s+n}\right)^2 
\right]^{-1}
\nonumber \\
&= \frac{N}{\Delta},
\end{align}
where $\Delta$ is again given by Eq.(S5) and
\begin{align}
N=
{(n+1)^2 (n-s)^2 - 2x (n^2-1)(n^2-s^2)\cos{\phi}
+x^2 (n-1)^2 (n+s)^2}.
\end{align}

The back transmittance and reflectance are obtained from the Fresnel formulae
for the substrate-air interface:
\begin{align}
R_b=\left(\frac{s-1}{s+1}\right)^2, \qquad T_b=\frac{4s}{(s+1)^2}.
\end{align}
We are now in position to evaluate total transmittance using Eq.(1),
with $\alpha=0$. We have
\begin{align}
 {\cal T}&=\frac{T_f T_b}{1-\tilde{R}_f R_b}
=
 \frac{16 n^2 s x}{\Delta} \frac{4s}{(s+1)^2}
\left[
1- \frac{(s-1)^2}{(s+1)^2} \frac{N}{\Delta}
\right]^{-1}
\nonumber \\
&=\frac{64 n^2 s^2 x}
{(s+1)^2 \Delta - (s-1)^2 N}.
\end{align}
Substituting $\Delta$ and $N$, the
denominator of this expression becomes
\begin{align}
&\quad\,\,(s+1)^2 [(n+1)^2 (n+s)^2 - 2x (n^2-1)(n^2-s^2)\cos{\phi}
+x^2 (n-1)^2 (n-s)^2]
\nonumber \\
&-(s-1)^2 [(n+1)^2 (n-s)^2 - 2x (n^2-1)(n^2-s^2)\cos{\phi}
+x^2 (n-1)^2 (n+s)^2].
\end{align}
We evaluate the powers of $x$ separately. First, for $x^0$:
\begin{align}
(s+1)^2 (n+1)^2 (n+s)^2 - (s-1)^2 (n+1)^2 (n-s)^2
&=4s(n+1)^3(n+s^2)
\nonumber \\
&=4s B,
\end{align}
for $x^1$:
\begin{align}
- 2x (n^2-1)(n^2-s^2)\cos{\phi} \left[(s+1)^2-(s-1)^2\right]
&=-8sx(n^2-1)(n^2-s^2) \cos{\phi}
\nonumber \\
&=-4s C x \cos{\phi}
\end{align}
and for $x^2$:
\begin{align}
x^2 (n-1)^2 
\left[(s+1)^2 (n-s)^2-(s-1)^2(n+s)^2 \right]
&=
4sx^2 (n-1)^3(n-s^2)
\nonumber \\
&=4s D x^2,
\end{align}
where the coefficients $B$, $C$ and $D$ are the same as in
Ref.\cite{swanepoel}. Putting these parts back together, and defining
$A=16n^2s$, Eq.(S11) becomes
\begin{align}
{\cal T}=\frac{Ax}{B-C x \cos{\phi} + D x^2},
\end{align}
which is Eq.(4) of Ref.\cite{swanepoel}.


\begin{thebibliography}{9}
\bibitem{thickness} The thicknesses are all given in terms of the
wavelength in the medium, that is $\lambda=\lambda_0/n$.
\end{thebibliography}
\end{document}